\documentclass[aps,pre,reprint,unsortedaddress,showpacs,amsmath,amssymb,twocolumn,preprintnumbers]{revtex4-1}

\usepackage{graphics,graphicx}

\begin{document}
\preprint{Physical Review E}

\title{Log-stable law of energy dissipation as a framework of turbulence intermittency}

\author{Hideaki Mouri}
\affiliation{Meteorological Research Institute, Nagamine, Tsukuba 305-0052, Japan}

\date{March 6, 2015}


\begin{abstract}
To describe the small-scale intermittency of turbulence, a self-similarity is assumed for the probability density function of a logarithm of the rate of energy dissipation smoothed over a length scale among those in the inertial range. The result is an extension of Kolmogorov's classical theory in 1941, i.e., a one-parameter framework where the logarithm obeys some stable distribution. Scaling laws are obtained for the dissipation rate and for the two-point velocity difference. They are consistent with theoretical constraints and with the observed scaling laws. Also discussed is the physics that determines the value of the parameter.
\end{abstract}


\maketitle


\section{Introduction} \label{S1}

Consider a fully developed state of three-dimensional homogeneous and isotropic turbulence that occurs in an incompressible fluid. The kinetic energy is supplied at a large length scale and is dissipated into heat by viscosity $\nu$ at around a small scale of the Kolmogorov length $\eta_{{\rm K}}$ \cite{ko41}.

Between these two, there is an inertial range of length scales, where the kinetic energy is transferred on average to the smaller and smaller scales without essentially any dissipation. This range has no characteristic scales about its lengths $r$ and about random variables defined at each of the lengths $r$. For the spatial average $\langle \cdot \rangle$ of such a variable $X_r$, we expect the scale invariance $\langle X_{\gamma r} \rangle = C_{\gamma} \langle X_r \rangle$ with $C_{\gamma} > 0$ that is a function of $\gamma > 0$. The solution is a power law $\langle X_r \rangle \varpropto r^{\tau}$, where the scaling exponent $\tau$ is unlikely to depend on the details of the energy supply \cite{ko41}. Below the inertial range, there is a dissipation range. The power-law scaling does not persist because of the existence of the Kolmogorov length $\eta_{\rm K}$ as a characteristic scale \cite{vn49}.

Turbulence is intermittent at these length scales in the inertial and dissipation ranges. An example is the rate of energy dissipation per unit mass $\varepsilon = \nu \sum_{i,j =1}^3 (\partial v_i /\partial {x_j}+ \partial v_j /\partial {x_i})^2 /2$, where $v_i$ is the velocity in the direction of the coordinate $x_i$. This rate is uniform according to the 1941 theory of Kolmogorov \cite{ko41}, but actually it fluctuates in space. If smoothed over a scale in the inertial or dissipation range (see Sec.~\ref{S2}), it is significant only within a fraction of the space. The fraction decreases with a decrease in the scale \cite{bt49,gsm62}. Such small-scale intermittency has been intensively studied for decades \cite{igk09}.

The central issue is to formulate a statistical framework that describes the rate $\varepsilon$ of energy dissipation. Although the dissipation itself is negligible in the inertial range, a power-law scaling is expected for the rate $\varepsilon_r$ smoothed over length scale $r$ in this range,
\begin{subequations}
\label{eq1}
\begin{equation}
\label{eq1a}
\left\langle \varepsilon_r^m \right\rangle \varpropto r^{\tau_m}.
\end{equation}
Since the dissipation rate is positive as $\varepsilon > 0$, the exponent $\tau_m$ has to satisfy an inequality explained in Sec.~\ref{S2} for the dimension  $D = 1$, $2$, or $3$ of the smoothing region \cite{n69},
\begin{equation}
\label{eq1b}
\frac{\tau_m}{m} > -D \ \ \ \mbox{at} \ m > 0.
\end{equation}
This does not apply to the rate of energy transfer through the scale $r$, which takes both positive and negative values \cite{igk09}. Also from an assumption for the intermittency (see Sec.~\ref{S2}), we expect an asymptote \cite{n94},
\begin{equation}
\label{eq1c}
\frac{\tau_m}{m} \rightarrow -D
\ \ \
\mbox{as}
\
m \rightarrow +\infty .
\end{equation}
\end{subequations}
The framework has to be justified by mathematical consistency with these and other constraints \cite{wf00}. We would be left with free parameters that are to be determined by physics outside the framework.

Kolmogorov in 1962 \cite{ko62} was the first to propose a theory for the small-scale intermittency, by assuming that $\varepsilon_r$ obeys a log-normal distribution or equivalently $\ln \varepsilon_r$ obeys a Gaussian (normal) distribution throughout scales $r$ in the inertial range. However, since the resultant exponent $\tau_m$ does not satisfy Eq.~(\ref{eq1b}), this log-normal theory does not deserve to be a statistical framework of the dissipation rate.

The log-normal theory is nevertheless of interest because the probability density function (PDF) of $\ln \varepsilon_r$ is self-similar among those scales $r$. While its position and its width are different, its shape remains identical. To see mathematics of this self-similarity, we ignore fluctuations at the largest scale $R$ of the inertial range and divide the scales as $r = r_1 < r_2 < \cdots < r_{N+1} = R$,
\begin{equation}
\label{eq2}
\ln \left( \frac{\varepsilon_r}{\varepsilon_R} \right) = \sum_{n=1}^{N} \, \ln \left( \frac{\varepsilon_{r_n}}{\varepsilon_{r_{n+1}}} \right) .
\end{equation}
The summands $\ln ( \varepsilon_{r_n} / \varepsilon_{r_{n+1}} )$ are set to be Gaussian and to be independent of one another. Since the Gaussianity is invariant under addition of its independent variables, the sum $\ln (\varepsilon_r / \varepsilon_R)$ is again Gaussian. The self-similarity in such a sense, if actual, should be useful to formulating the framework.

Gaussian distributions make up a two-parameter family, which is included in a four-parameter family of {\it stable} distributions \cite{f71,m82,s13}. If and only if the distribution is stable, the shape of its PDF is invariant under addition of its variables. For scales $r$ in the inertial range, Kida \cite{k91} considered that $\ln (\varepsilon_r / \varepsilon_R)$ obeys a stable distribution. Two of its four parameters were determined mathematically, while the other two were left as free parameters. This log-stable theory is necessary and sufficient for a self-similar PDF to describe the fluctuations of $\ln (\varepsilon_r / \varepsilon_R)$. However, the theory as a whole is not consistent with Eq.~(\ref{eq1b}). In fact, its special case is the log-normal theory.

The same self-similarity of $\ln (\varepsilon_r / \varepsilon_R)$ exists in the 1941 theory of Kolmogorov \cite{ko41}, i.e., $\tau_m = 0$, which is not intermittent but is consistent with Eq.~(\ref{eq1b}) and is regarded as the {\it degenerate} case of some stable distribution. We search for an extension of this classical theory within the whole family of the stable distributions. The result is a one-parameter framework where $\ln (\varepsilon_r / \varepsilon_R)$ in the inertial range has a self-similar PDF and is consistent with Eq. (\ref{eq1b}) and also with Eq.~(\ref{eq1c}). We discuss the application to the velocity field, consistency with the actual turbulence, physics that determines the parameter value, and differences from the existing frameworks.

\section{Basic Settings} \label{S2}

The dissipation rate $\varepsilon$ is smoothed over length scales $r$ in the inertial range. Since its boundaries are not definite, it is set to be wide enough as compared with ranges of scales that could be used as its largest and smallest scales. The largest scale is fixed at $R$, which is well below the scale of the energy supply \cite{ko41}. We do not fix the smallest scale because the inertial range could extend to any small scale $r > \eta_{\rm K}$ in accordance with the Reynolds number of the turbulence.

\begin{figure}[tbp]
\rotatebox{90}{
\resizebox{4.cm}{!}{\includegraphics*[6.1cm,3.7cm][16.9cm,24.7cm]{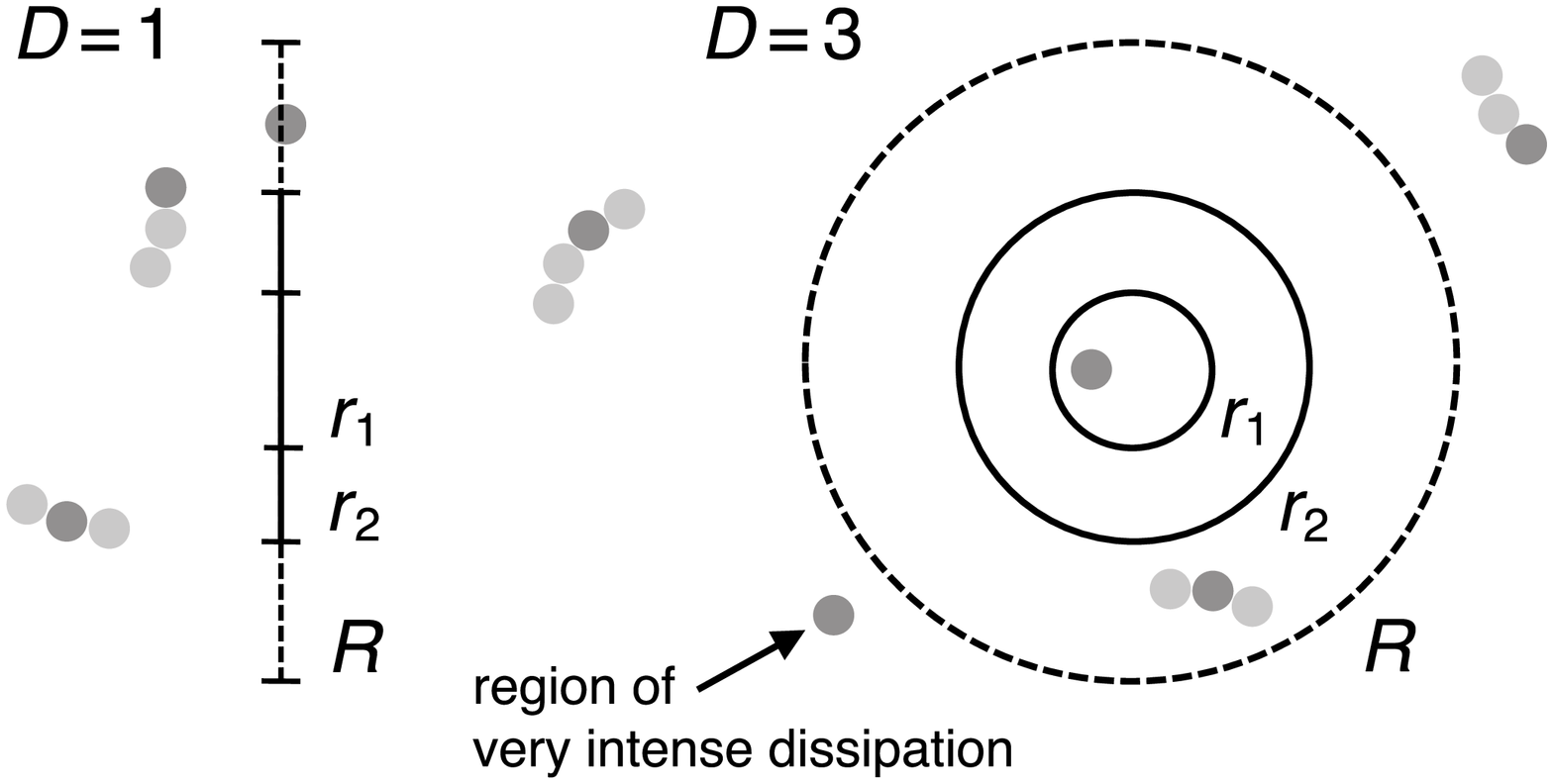}}
}
\caption{\label{f1} Schematic illustration of smoothing regions for $D=1$ in Eq.~(\ref{eq3a}) and for $D=3$ in Eq.~(\ref{eq4a}) on a two-dimensional cut of the space. The grey and dark grey areas denote regions of intense dissipation and of very intense dissipation.}
\end{figure}

We consider the one-dimensional smoothing case $D=1$. As illustrated in Fig.~\ref{f1}, the rate $\varepsilon$ is averaged over a segment of length $r$ centered at a position $x$ along a line in the three-dimensional space,
\begin{subequations}
\label{eq3}
\begin{equation}
\label{eq3a}
\varepsilon_r(x) = \frac{1}{r} \int_{\vert x^{\prime}-x \vert \le r/2} \varepsilon (x^{\prime}) dx^{\prime}.
\end{equation}
The direction of this line is fixed arbitrarily in the isotrop-ic turbulence. At the same position $x$, we define a random variable $\chi_r$ as
\begin{equation}
\label{eq3b}
\chi_r(x) = \ln \left[ \frac{r \varepsilon_r(x)}{R \varepsilon_R(x)} \right] \le 0 \ \ \ \mbox{at} \ r \le R.
\end{equation}
\end{subequations}
The inequality $\chi_r \le 0$ is due to the positivity $\varepsilon > 0$ in Eq.~(\ref{eq3a}). We also consider the three-dimensional smoothing case $D=3$, by averaging the rate $\varepsilon$ over a spherical volume of diameter $r$ centered at a position {\boldmath $x$},
\begin{subequations}
\label{eq4}
\begin{equation}
\label{eq4a}
\varepsilon_r(\mbox{\boldmath $x$}) 
= \frac{6}{\pi r^3} \int_{\vert \mbox{\boldmath $x$}^{\prime}-\mbox{\boldmath $x$} \vert \le r/2} 
                    \varepsilon (\mbox{\boldmath $x$}^{\prime}) d\mbox{\boldmath $x$}^{\prime},
\end{equation}
and
\begin{equation}
\label{eq4b}
\chi_r(\mbox{\boldmath $x$}) 
= \ln \left[ \frac{r^3 \varepsilon_r(\mbox{\boldmath $x$})}{R^3 \varepsilon_R(\mbox{\boldmath $x$})} \right] \le 0 
\ \ \ 
\mbox{at} \ r \le R.
\end{equation}
\end{subequations}
The center of the smoothing region is fixed at a position in the space so as to regard $\varepsilon_r$ and $\chi_r$ as random functions of $r$ in the inertial range up to the largest scale $R$. Their spatial averages are obtained by shifting the position of that center.

The smaller smoothing region is included in the larger so long as their centers are common. Also since the rate $\varepsilon$ is positive, its integral over the region $\varpropto r^D \varepsilon_r$ is increasing in the scale $r$. The moment $\langle (r^D \varepsilon_r)^m \rangle \varpropto r^{Dm+\tau_m}$ at $m > 0$ is increasing as well, i.e., $Dm+\tau_m > 0$ \cite{n69}. This corresponds to Eq.~(\ref{eq1b}), which always holds regardless of relative positions of the smoothing regions because the turbulence is homogeneous.

Another constraint is expected from the intermittency \cite{n94}. We assume that $r^D \varepsilon_r / R^D \varepsilon_R$ is independent of $R^D \varepsilon_R$ and takes any value from $0$ to $1$. Then $r^D \varepsilon_r$ and $R^D \varepsilon_R$ take the same maximum value, i.e., $\langle (r^D \varepsilon_r)^m \rangle^{1/m} \rightarrow \langle (R^D \varepsilon_R)^m \rangle^{1/m}$ as $m \rightarrow +\infty$. It means $r^{(Dm+\tau_m)/m} \rightarrow r^0$ and hence Eq.~(\ref{eq1c}). The maximum values are the same even at $r \ll R$. That is, any very intense dissipation has been assumed to occur in an isolated point-like region (see Fig.~\ref{f1}). Although the energy dissipation is intense in organized structures, the intensity within each of the structures is not uniform but is enhanced in its point-like regions \cite{igk09}. They are at $r>0$, so that they could exist in smoothing regions of any dimension $D$ and hence Eq. (\ref{eq1c}) holds for any value of $D \le 3$.

\section{Formulation} \label{S3}

The following conditions are imposed on the random variable $\chi_r$ in the inertial range at $r \le R$: (i) $\chi_{r_1} - \chi_{r_2}$ has a PDF that depends only on $r_1/r_2$ for any pair of $r_1$ and $r_2$; (ii) $\chi_{r_1} - \chi_{r_2}$, $\chi_{r_2} - \chi_{r_3}$, ...,  $\chi_{r_{N-1}} - \chi_{r_N}$ do not depend on one another for any finite series of $r_1 < r_2 < \cdots < r_N$; and (iii) $\chi_r$ has a self-similar PDF. If the sign $\triangleq$ is used to denote that the two random variables have the same PDF, the condition (iii) is such that a constant $C_{r_1,r_2} > 0$ is found for any pair of $r_1$ and $r_2$ to have $C_{r_1,r_2} \chi_{r_1} \triangleq \chi_{r_2}$. From these three conditions, it follows that $\chi_r$ is {\it strictly} stable at each of the scales $r$. The above self-similarity is described as
\begin{equation}
\label{eq5}
\chi_r \triangleq \left[ \ln \left( R^{\beta}/r^{\beta} \right) \right]^{1/\alpha} \chi_{\ast} 
\ \ \ 
\mbox{with} \ 0 < \alpha \le 2 \ \mbox{and} \ \beta > 0 .
\end{equation}
Here $\chi_{\ast}$ is identical to $\chi_r$ at $\ln (R^{\beta}/r^{\beta}) = 1$. We are also to use an auxiliary condition that $\varepsilon_R$ is uniform in space, i.e., $\varepsilon_R = \langle \varepsilon \rangle$, although it is only for simplicity because $\varepsilon_R$ does not affect the PDF of $\chi_r$ nor of $\varepsilon_r / \varepsilon_R$ at least under the condition (i).

Here is a mathematical explanation \cite{f71,s13}. The conditions (i) and (ii) permit us to regard $\chi_r$ as a stochastic L\'evy process for the time parameter $t = \ln (R^{\beta}/r^{\beta}) \ge 0$. Any process $\chi_t$ at $t \ge 0$ is a L\'evy process if $\chi_t = 0$ at $t=0$, if $\chi_t$ is a continuous function of $t$, if $\chi_{t_1} - \chi_{t_2}$ has a PDF that depends only on $t_1 - t_2$ for any pair of $t_1$ and $t_2$, and if $\chi_{t_1} - \chi_{t_2}$, $\chi_{t_2} - \chi_{t_3}$, ...,  $\chi_{t_{N-1}} - \chi_{t_N}$ do not depend on one another for any finite series of $t_1 < t_2 < \cdots < t_N$. The parameter $\beta$ is to define the scale $r$ that corresponds to the time $t = 1$.

The L\'evy process $\chi_t$ is stable if constants $C_{t_1,t_2}^{\ } >0$ and $C_{t_1,t_2}^{\, \prime}$ are found for any pair of $t_1$ and $t_2$ to have $C_{t_1,t_2}^{\ } \chi_{t_1}^{\ }+C_{t_1,t_2}^{\, \prime} \triangleq \chi_{t_2}^{ }$. If all the pairs yield $C_{t_1,t_2}^{\, \prime}=0$, the stability is said to be strict. This is our condition (iii). The constant $C_{t_1,t_2}^{\, \prime}$ serves as a shift of the PDF. It is not required for $\chi_t$ used here, judging from our discussion in Sec.~\ref{S2}.

By defining $\chi_{\ast}$ as $\chi_t$ at $t=1$, the strictly stable process is described as $\chi_t \triangleq t^{1/\alpha} \chi_{\ast}$ with a parameter $0 < \alpha \le 2$. This is Eq.~(\ref{eq5}). At each $t$, the distribution of $\chi_t$ is strictly stable with the same parameter $\alpha$. To be specific, if $\chi_{t,1}^{\, \prime}$ and $\chi_{t,2}^{\, \prime}$ are independent copies of $\chi_t^{\ }$, we have $C_{1}^{\ } \chi_{t,1}^{\, \prime} + C_2^{\ } \chi_{t,2}^{\, \prime} \triangleq (C_{1}^{\alpha}+C_{2}^{\alpha})^{1/\alpha} \chi_t^{\ }$ for any pair of constants $C_1^{\ } > 0$ and $C_2^{\ } > 0$. The shape of the PDF is invariant without its shift.

The condition (i) means homogeneity of $\ln (\varepsilon_{r_n} / \varepsilon_{r_{n+1}})$ throughout scales $r$ in the inertial range, while the condition (ii) means their independence. Under these two, no characteristic scale is permissible, so that we expect the existence of the power-law scaling $\langle \varepsilon_r^m \rangle \varpropto r^{\tau_m}$ (see Sec. \ref{S1}). They are actually idealized conditions. For example, $\ln (\varepsilon_{r_1}/\varepsilon_{r_2})$ and $\ln (\varepsilon_{r_2}/\varepsilon_{r_3})$ are not independent of each other for narrowly separated scales, say, $r_3/r_1 \lesssim 10$ \cite{pnp96}, although such narrow separations are not important if the inertial range is wide enough.

The condition (iii) means a self-similarity of the PDF of $\ln (\varepsilon_r / \varepsilon_R)$, albeit requiring its shift, among scales $r$ in the inertial range. If the shape of such a self-similar PDF is used to define the intermittency, it is not dependent on $r$. The dependence does exist for the intermittency of $\varepsilon_r / \varepsilon_R$ because the PDF is no longer self-similar. From Eq. (\ref{eq5}) via Eq.~(\ref{eq3b}) or (\ref{eq4b}), we find that $\varepsilon_r  / \varepsilon_R$ is increasingly intermittent with a decrease in the scale $r$ from the largest scale $R$.

Stable distributions are described generally by four parameters, but those in the strict sense make up a family of three parameters, where excluded is a parameter to shift the PDF \cite{f71,s13}. Except for $\alpha = 1$, the characteristic function of the distribution of $\chi_{\ast}$ is
\begin{equation}
\label{eq6}
\left\langle \exp (i \chi_{\ast} \xi ) \right\rangle
=
\exp \left( -\lambda \vert \xi \vert^{\alpha} e^{i \pi \theta \xi / 2 \vert \xi \vert} \right) . 
\end{equation}
While $\lambda > 0$ determines the width of the PDF, $\alpha$ and $\theta$ determine its shape ($\vert \theta \vert \le \alpha$ for $0 < \alpha < 1$ and $\vert \theta \vert \le 2-\alpha$ for $1 < \alpha \le 2$). The stable distributions for $\alpha = 2$ and $\theta = 0$ are Gaussian.

We focus on the distributions for $0 < \alpha < 1$ and $\theta = \alpha$. They alone are totally skewed to the left, i.e., $\chi_{\ast} \le 0$ \cite{f71,s13}, which is required from Eq.~(\ref{eq3b}) or (\ref{eq4b}) via Eq. (\ref{eq5}). By using the above values of $\alpha$ and $\theta$ in the inverse Fourier transform of Eq.~(\ref{eq6}), the PDF $f(\chi_{\ast})$ is written as
\begin{equation}
\label{eq7}
f(\chi_{\ast})=
  \begin{cases}
  0 
  & 
  \mbox{at} \ \chi_{\ast} \ge 0 \\
  \displaystyle\sum_{n=1}^{\infty} 
  \frac{\lambda (-\lambda)^{n-1} \Gamma (n\alpha +1) \sin (\pi n \alpha) }{\pi n! \vert \chi_{\ast} \vert^{n\alpha +1}}
  &
  \mbox{at} \ \chi_{\ast} < 0.
  \end{cases}
\end{equation}
Here $\Gamma$ is the gamma function. The tail for $\chi_{\ast} \rightarrow -\infty$ is in the form of power law $\varpropto \vert \chi_{\ast} \vert^{-(\alpha+1)}$, which leads to divergence of moments such as the average $\langle \chi_{\ast} \rangle$.

To formulate the one-dimensional smoothing case $D=1$, we use Eq.~(\ref{eq3b}) to have $\langle r^m \varepsilon_r^m / R^m \varepsilon_R^m \rangle = \langle \exp (m \chi_r) \rangle$. Then $\langle \exp (m \chi_r) \rangle$ is obtained by substituting Eq.~(\ref{eq5}) into Eq.~(\ref{eq6}) and by replacing $\xi$ with $-im$. This replacement corresponds to an analytical continuation from $\xi$ to $-im$. It is justified at $m \ge 0$ by the convergence of $\langle \exp (m \chi_{\ast}) \rangle$, which is in turn justified by the form of $f(\chi_{\ast})$ in Eq.~(\ref{eq7}). The result is
\begin{equation}
\label{eq8}
\left\langle \frac{\varepsilon_r^m}{\varepsilon_R^m} \right\rangle
=
\left( r/R \right)^{-m+m^{\alpha} \beta \lambda \exp ( \pi \alpha /2 ) }
\ \ \
\mbox{at}
\
m \ge 0.
\end{equation}
However, $\langle \varepsilon_r^m / \varepsilon_R^m \rangle$ does not exist at $m<0$. The reason is the power-law tail of $f(\chi_{\ast})$ and the resultant divergence of $\langle \exp (m \chi_{\ast}) \rangle$ at $m<0$. Negative-order moments observed in numerical simulations and in experiments are attributable to, e.g., random noise that is inherent in any actual data.

Finally, the parameters $\beta$ and $\lambda$ are eliminated by making use of the auxiliary condition $\varepsilon_R = \langle \varepsilon \rangle$.  The left-hand side of Eq.~(\ref{eq8}) is reduced to $\langle \varepsilon_r^m \rangle / \langle \varepsilon \rangle^m$. At $m=1$, it is $\langle \varepsilon_r \rangle / \langle \varepsilon \rangle = 1$. The corresponding exponent for $r/R$ in the right-hand side is $-1+ \beta \lambda \exp ( \pi \alpha /2 ) = 0$. Accordingly,
\begin{subequations}
\label{eq9}
\begin{equation}
\label{eq9a}
\frac{\left\langle \varepsilon_r^m\right\rangle }{\langle \varepsilon \rangle^m} = \left( r/R \right)^{-m+m^{\alpha}},
\end{equation}
and hence as the exponent $\tau_m$ for $D=1$,
\begin{equation}
\label{eq9b}
\tau_m = -m+m^{\alpha}
\ \ \
\mbox{with}
\
0 < \alpha < 1
\
\mbox{at}
\
m \ge 0.
\end{equation}
\end{subequations}
Even if $\varepsilon_R$ is not uniform, $\varepsilon_R$ does not affect $\varepsilon_r / \varepsilon_R$ under the condition (i). We have $\langle \varepsilon_r^m / \varepsilon_R^m \rangle \langle \varepsilon_R^m \rangle = \langle \varepsilon_r^m \rangle$, through which Eq.~(\ref{eq8}) leads to $\langle \varepsilon_r^m \rangle \varpropto r^{\tau_m}$ with the same Eq.~(\ref{eq9b}). Then, Eq.~(\ref{eq9a}) could be regarded as a result conditioned on a given value $\langle \varepsilon \rangle$ of the fluctuating rate $\varepsilon_R$ \cite{gsm62}.

This is the necessary and sufficient framework for the case of $D=1$ where $\ln (\varepsilon_r / \varepsilon_R )$ has a self-similar PDF and satisfies Eq.~(\ref{eq1b}). Also satisfied is Eq.~(\ref{eq1c}). The intermittency is described by the parameter $\alpha$. Especially in the limit $\alpha \rightarrow 1$,  we reproduce the 1941 theory of Kolmogorov \cite{ko41}, i.e., $\tau_m = 0$, for which Eq.~(\ref{eq1c}) does not hold because the intermittency does not persist.

The three-dimensional smoothing case $D=3$ is also formulated, by replacing the length ratio $r/R$ in Eq.~(\ref{eq9a}) with the volume ratio $r^3/R^3$,
\begin{subequations}
\label{eq10}
\begin{equation}
\label{eq10a}
\frac{\left\langle \varepsilon_r^m\right\rangle }{\langle \varepsilon \rangle^m} 
= \left( r^3/R^3 \right)^{-m+m^{\alpha}}
= \left( r  /R   \right)^{-3m+3m^{\alpha}},
\end{equation}
and hence as the exponent $\tau_m$ for $D=3$,
\begin{equation}
\label{eq10b}
\tau_m = -3m+3m^{\alpha}
\ \ \
\mbox{with}
\
0 < \alpha < 1
\
\mbox{at}
\
m \ge 0.
\end{equation}
\end{subequations}
Between $D=1$ and $D=3$, the exponent $\tau_m$ is different. We have $\tau_m = -Dm+Dm^{\alpha}$, which holds for $D=2$ as well. The value of the parameter $\alpha$ could depend on the smoothing dimension $D$, although there has been found no mathematical relation.

The present formulation is almost equivalent to that of Kida \cite{k91}. However, his stable distribution for $\chi_r$ is not in the present three-parameter strict sense but is in the four-parameter general sense. The resultant exponent has the form of $\tau_m = -\mu (m^{\alpha}-m)/(2^{\alpha}-2)$ with two free parameters $\mu  > 0$ and $0 < \alpha \le 2$, even if the smoothing dimension $D$ is given. We have imposed an additional constraint of $\chi_r \le 0$ to eliminate the parameter $\mu$ and to obtain Eqs.~(\ref{eq9b}) and (\ref{eq10b}) where $\tau_m$ always satisfies the inequality of Eq.~(\ref{eq1b}).

\section{Application to Velocity Field} \label{S4}

The power-law scaling in the inertial range is expected also for the two-point velocity difference $\delta u_r(x) = u(x+r)-u(x)$,
\begin{subequations}
\label{eq11}
\begin{equation}
\label{eq11a}
\left\langle \vert \delta u_r^m \vert \right\rangle \varpropto r^{\zeta_m}
\ \ \
\mbox{at}
\
m \ge 0 .
\end{equation}
Here $u$ is longitudinal or lateral, i.e., parallel or perpendicular to the line through the two points. The coordinate $x$ is along this line. We study $\vert \delta u_r \vert$ instead of $\delta u_r$ itself \cite{ko62}. Since $\delta u_r$ could take the value of $0$, the moment $\langle \vert \delta u_r^m \vert \rangle$ does not exist at $m < 0$.

If the velocity $u$ is to be bounded such that the flow does not become supersonic and hence does not become compressible, there is a necessary condition
\begin{equation}
\label{eq11b}
\frac{d \zeta_m}{dm} \ge 0 .
\end{equation}
\end{subequations}
This is because $\langle \vert \delta u_r^{m+m^{\prime}} \vert \rangle / \langle \vert \delta u_r^m \vert \rangle \varpropto r^{\zeta_{m+m^{\prime}} - \zeta_m}$ has to be bounded for $m^{\prime} > 0$ even if the inertial range extends to any small scale, i.e., $\zeta_{m+m^{\prime}} - \zeta_m \ge 0$ \cite{f91}.

The exponent $\zeta_m$ is obtained from the exponent $\tau_m$ for the dissipation rate $\varepsilon_r$. Based on a velocity scale $(r \varepsilon_r)^{1/3}$, there is a relation known as Kolmogorov's refined similarity hypothesis \cite{ko62},
\begin{equation}
\label{eq12}
\langle \vert \delta u_r^m \vert \rangle \varpropto \langle  (r \varepsilon_r)^{m/3} \rangle 
\ \ \
\mbox{and}
\ \ \
\zeta_m = \tau_{m/3} + \frac{m}{3} .
\end{equation}
To the longitudinal velocity, Eq.~(\ref{eq12}) is plausible because the energy dissipation is intense in regions under intense strain. We do not use the lateral velocity. Its two-point difference is rotational and is not related closely to the rate of the dissipation \cite{m07}.

We consider the one-dimensional smoothing case $D=1$. The two points for the velocity difference $\delta u_r$ are the ends of the line segment used to smooth the dissipation rate $\varepsilon_r$. From $\tau_m = -m+m^{\alpha}$ in Eq.~(\ref{eq9b}) via Eq.~(\ref{eq12}),
\begin{equation}
\label{eq13}
\zeta_m = \left( \frac{m}{3} \right)^{\alpha}
\ \ \
\mbox{with}
\
0 < \alpha < 1
\
\mbox{at}
\
m \ge 0.
\end{equation}
The exponent $\zeta_m$ satisfies Eq.~(\ref{eq11b}) and other constraints such as $\zeta_2 \ge 2/3$ \cite{cf94}. In the limit $\alpha \rightarrow 1$, we reproduce the 1941 theory of Kolmogorov \cite{ko41}, i.e., $\zeta_m = m/3$.

We also consider the three-dimensional smoothing case $D=3$. The two points for the velocity difference $\delta u_r$ are poles of the sphere used to smooth the dissipation rate $\varepsilon_r$. From $\tau_m = -3m+3m^{\alpha}$ in Eq.~(\ref{eq10b}) via Eq.~(\ref{eq12}), we have $\zeta_m = -2m/3+3(m/3)^{\alpha}$. As $m \rightarrow +\infty$, the exponent $\zeta_m$ becomes inconsistent with Eq.~(\ref{eq11b}) and becomes negative. The reason is $\tau_m /m \rightarrow -3$ that satisfies Eq.~(\ref{eq1c}). Thus, Eqs.~(\ref{eq1c}), (\ref{eq11b}), and (\ref{eq12}) are inconsistent with one another in the case of $D=3$.

The dissipation exponent $\tau_m$ is different between $D=1$ and $D=3$, while the velocity exponent $\zeta_m$ is unique. It is hence possible that the relation of $\tau_m$ to $\zeta_m$ in Eq.~(\ref{eq12}) is suited to $D=1$ but is not to $D=3$. The velocity difference $\delta u_r$ is an integral of the velocity derivative $\partial u/ \partial x$ over the separation $r$, while an integral of $(\partial u/ \partial x)^2$ is one of the components of the dissipation rate $\varepsilon_r$. Regions of these two integrations are identical in the case of $D=1$. They are not identical in the case of $D=3$, for which it would be of interest to reconsider whether or not any relation exists between the velocity difference $\delta u_r$ and the dissipation rate $\varepsilon_r$.

\begin{figure}[tbp]
\resizebox{8.0cm}{!}{\includegraphics*[3.5cm,9.7cm][17.0cm,27.0cm]{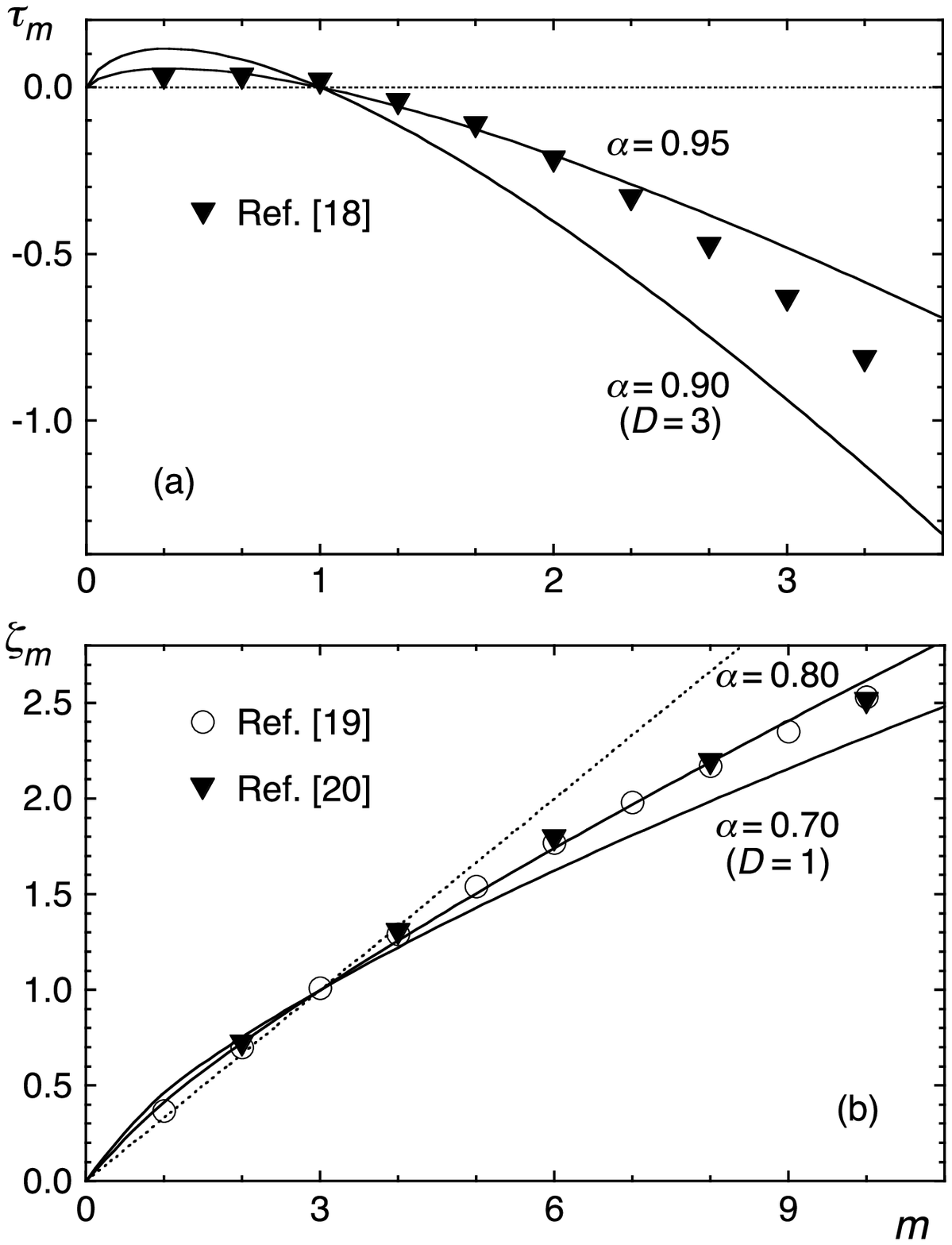}}
\caption{\label{f2} Comparison of our framework with numerical simulations of forced steady states of homogeneous and isotropic turbulence at the Reynolds number for the Taylor microscale of $216$ \cite{c97}, $460$ \cite{g02}, and $600$ \cite{b10}: (a) dissipation exponent $\tau_m$ for $D=3$ in Eq.~(\ref{eq10b}), and (b) velocity exponent $\zeta_m$ for $D=1$ in Eq.~(\ref{eq13}). The dotted lines denote the 1941 theory of Kolmogorov \cite{ko41}.}
\end{figure} 

\section{Comparison with Observations} \label{S5}

Figure \ref{f2} compares the scaling exponents of our framework with those observed in numerical simulations of homogeneous and isotropic turbulence \cite{c97,g02,b10}. Since our framework is well consistent with the simulations, we conclude that our conditions (i)--(iii) used in Sec.~\ref{S3} are at least good approximations of the actual turbulence. This conclusion applies especially to the condition (iii), which says that $\ln (\varepsilon_r / \varepsilon_R)$ has a self-similar PDF.

The three-dimensional smoothing case $D=3$ has been studied for the dissipation exponent $\tau_m$ in Eq.~(\ref{eq10b}). It yields $\alpha \simeq 0.90$--$0.95$. On the other hand, in the one-dimensional smoothing case $D=1$, the data of $\tau_m$ are not available, except for a surrogate of the dissipation rate $\varpropto (\partial u/ \partial x)^2$ that tends to be more intermittent \cite{w96}. We have instead used the velocity exponent $\zeta_m$ in Eq.~(\ref{eq13}). It yields $\alpha \simeq 0.7$--$0.8$. Thus $\alpha$ takes different values between the cases of $D=1$ and $D=3$.

With a numerical simulation or with an experiment, we would have to compare our framework again in the future. The conditions (i) and (ii), i.e., homogeneity and independence of $\ln (\varepsilon_{r_n} / \varepsilon_{r_{n+1}})$ in the inertial range, would hold if the Reynolds number were high enough to set up a wide inertial range. This is not the case in the existing simulations and experiments \cite{igk09,ab06}. We expect that the future data are at a high enough Reynolds number, which would also permit the more accurate observations of $\tau_m$ and $\zeta_m$. The auxiliary condition, i.e., uniformity of $\varepsilon_R$, also does not hold because $\varepsilon_R$ fluctuates significantly even if the turbulence is fully developed and is filling the space \cite{ll59,m09}. Although such fluctuations are unlikely to affect the power law in the form of $\langle \varepsilon_r^m \rangle \varpropto r^{\tau_m}$, we expect that the future data are of enough size to study the scaling of $\langle \varepsilon_r^m \rangle$ at each value of $\varepsilon_R$. Then, our framework should exhibit a better consistency with the simulation or with the experiment if our self-similarity condition (iii) holds for the actual turbulence.

\section{Explanation for Parameter Value} \label{S6}

Thus far, we have discussed mathematics of the dissipation rate $\varepsilon_r$ to formulate a statistical framework of its log-stable law. If this framework does apply to the small-scale intermittency, the observed value of the parameter $\alpha$ is explainable by its physics. Here is an attempt of such an explanation, which is based on the transition from the inertial range to the dissipation range in the one-dimensional smoothing case $D=1$.

The transition is determined by the Reynolds number $\mbox{Re}_r$. For a region of length $r$ centered at some position $x$, we follow Kolmogorov \cite{ko62} to define $\mbox{Re}_r$ by making use of a velocity scale $(r \varepsilon_r)^{1/3}$,
\begin{subequations}
\label{eq14}
\begin{equation}
\label{eq14a}
\mbox{Re}_r(x) = \frac{r [r \varepsilon_r(x)]^{1/3}}{\nu}.
\end{equation}
The same velocity scale exists in Eq.~(\ref{eq12}). From $\mbox{Re}_{\eta} = 1$, the transition length scale $\eta$ is defined as
\begin{equation}
\label{eq14b}
\eta(x) = \frac{\nu^{3/4}}{\varepsilon_{\eta}^{1/4}(x)}.
\end{equation}
\end{subequations}
The scale $\eta$ fluctuates spatially with fluctuations of the dissipation rate $\varepsilon_{\eta}$ \cite{ko62}. Since $\langle \eta^{-4m} \rangle$ is related to $\langle \varepsilon_{\eta}^m \rangle$, we assume that $\langle \varepsilon_r^m \rangle$ in the inertial range transits to $\langle \varepsilon_r^m \rangle$ in the dissipation range at the length of $r = C_{\ast} \langle \eta^{-4m} \rangle^{-1/4m}$ \cite{ys05}. The constant $C_{\ast} \gtrsim 10$ is due to the Reynolds number $\mbox{Re}_r = C_{\ast}^{4/3}$ for the actual occurrence of the transition \cite{igk09}.

The inertial range exhibits $\langle \varepsilon_r^m \rangle / \langle \varepsilon \rangle^m = (r/R)^{\tau_m}$ if we use Eq.~(\ref{eq9a}), which yields
\begin{subequations}
\label{eq15}
\begin{equation}
\label{eq15a}
\frac{d}{dm} \ln \left( \frac{ \langle \varepsilon_r^m \rangle }{\langle \varepsilon \rangle ^m} \right)
=
\frac{d \tau_m}{dm} \ln ( r/R ) .
\end{equation}
Here $d \tau_m /dm = \alpha /m^{1-\alpha}-1 > 0$ at $m < m_{\ast} = \alpha^{1/(1-\alpha)}$ and $ d \tau_m /dm$ $\le 0$ at $m \ge m_{\ast}$. The value of $m_{\ast}$ is found at $0.3$--$0.4$ in the case of $\alpha \ge 0.5$.

The dissipation range exhibits smooth velocities such that $\partial v_i/ \partial x_j \simeq \mbox{constant}$ in any region of length $r \lesssim C_{\ast} \eta$, from which we have $\varepsilon_r(x) \simeq \varepsilon_{\eta}(x)$ and $\langle \varepsilon_r^m \rangle \simeq \langle \varepsilon_{\eta}^m \rangle$. Also by using the Kolmogorov length $\eta_{\rm K} = \nu^{3/4}/ \langle \varepsilon \rangle^{1/4}$ \cite{ko41},
\begin{equation}
\label{eq15b}
\frac{d}{dm} \ln \left( \frac{ \langle \varepsilon_r^m \rangle }{\langle \varepsilon \rangle ^m} \right) 
\simeq 
\frac{1}{m} \ln \left( \frac{ \langle \varepsilon_r^m \rangle }{\langle \varepsilon \rangle ^m} \right) 
\simeq
\frac{1}{m} \ln \left\langle \frac{\eta_{\rm K}^{4m}}{\eta^{4m}} \right\rangle .
\end{equation}
\end{subequations}
The derivative has been approximated by a linear slope from $\ln (\langle \varepsilon_r^m \rangle / \langle \varepsilon \rangle ^m) =0$ at $m=0$. It holds if $m$ is small, say, $m < m_{\ast}$. If $m$ were not small, the slope would have to be taken from $\ln (\langle \varepsilon_r^m \rangle / \langle \varepsilon \rangle ^m) =0$ at $m=1$.

Finally, our assumption is used to equate Eq.~(\ref{eq15a}) with Eq.~(\ref{eq15b}) at $r = C_{\ast} \langle \eta^{-4m} \rangle^{-1/4m}$. The result for $m = 1/4$ is
\begin{equation}
\label{eq16}
\left\langle \frac{R}{C_{\ast} \eta} \right\rangle \simeq \left( \frac{R}{C_{\ast} \eta_{\rm K}} \right)^{\left. 4/(4+d\tau_m/dm) \right\vert_{m=1/4}} .
\end{equation}
Here $\langle R / C_{\ast} \eta \rangle$ serves as the width of the inertial range. The classical width $R / C_{\ast} \eta_{\rm K}$ is reproduced in the limit $\alpha \rightarrow 1$, corresponding to the 1941 theory of Kolmogorov \cite{ko41}.

Given the classical width of the inertial range $R / C_{\ast} \eta_{\rm K}$, which is fixed by conditions external to the turbulence, $\langle R/ C_{\ast}\eta \rangle$ is minimal at $\alpha = 0.72$. Since $\alpha \simeq 0.7$--$0.8$ is observed for $D=1$ in Fig.~\ref{f2}, we consider that the observed value of $\alpha$ corresponds to the minimum of the width of the inertial range $\langle R/C_{\ast} \eta \rangle$. Its inverse $\langle R/ C_{\ast} \eta \rangle^{-1}$ serves as the mean efficiency of the energy dissipation. While $\eta$ is large in regions where the kinetic energy is effectively dissipated into heat, $\eta$ is not large in the other regions. The average over all the regions is $\langle \eta^{-1} \rangle^{-1}$. Thus, a plausible explanation is that the value of $\alpha$ is determined so as to maximize the mean efficiency of the energy dissipation $\langle R/C_{\ast} \eta \rangle^{-1}$ and thereby to minimize the width of the inertial range $\langle R/ C_{\ast} \eta \rangle$.

This explanation has used our auxiliary condition $\varepsilon_R = \langle \varepsilon \rangle$ and the resultant Eq.~(\ref{eq9a}). Even if $\varepsilon_R$ is not uniform, the same explanation is possible by considering $R / C_{\ast} \eta_{\rm K}$ and $\langle R / C_{\ast} \eta \rangle$ for each value of $\varepsilon_R$ (see Sec.~\ref{S3}).

To study the three-dimensional smoothing case $D=3$, we have to consider $\langle R^3/(C_{\ast} \eta)^3 \rangle$ instead of $\langle R/C_{\ast} \eta \rangle$, i.e., $m=3/4$ instead of $m = 1/4$. However, even if Eq.~(\ref{eq15b}) is rewritten for $m > m_{\ast}$, it leads to an inconsistent result that $\langle R^3/(C_{\ast} \eta)^3 \rangle$ is not minimal in the range of $0 < \alpha < 1$. The reason is attributable to the velocity scale $(r \varepsilon_r)^{1/3}$ in Eq.~(\ref{eq14a}), which is unlikely to be suited to $D=3$ as discussed for Eq.~(\ref{eq12}) in Sec.~\ref{S4}. An alternative approach is required to explain the observed value of the parameter $\alpha$ and the relation to its value for $D=1$, if possible on the basis of some definition of the mean efficiency of the energy dissipation.

\section{Comparison with Other Theories} \label{S7}

For the small-scale intermittency, most of the existing theories are based on the conditions (i) and (ii) used also in Sec.~\ref{S3}. They yield $\langle \varepsilon_r^m / \varepsilon_R^m \rangle = \langle \varepsilon_{\gamma R}^m / \varepsilon_R^m \rangle^{\ln (r/R) / \ln \gamma}$ with $r \le \gamma R < R$, which is necessary and sufficient for the existence of a power law $\langle \varepsilon_r^m / \varepsilon_R^m \rangle = (r/R)^{\tau_m}$ \cite{n69}. It is
 equivalent to $\langle \varepsilon_r^m \rangle \varpropto r^{\tau_m}$ because $\varepsilon_r / \varepsilon_R$ does not depend on $\varepsilon_R$ under the condition (i). To constrain the exponent $\tau_m$, each theory has some additional condition.

The current paradigm is multifractality \cite{pf85}. To formulate this into a statistical framework \cite{m91,f95}, instead of using our self-similarity condition (iii), the {\it large} deviation theory is used for the sum of $\ln (\varepsilon_{r_n} / \varepsilon_{r_{n+1}})$ in Eq.~(\ref{eq2}) with $r_n / r_{n+1} = \mbox{constant}$, i.e., with $N \varpropto -\ln (r/R)$. This limit theorem says that the PDF of $\ln (\varepsilon_r / \varepsilon_R) / \ln (r/R)$ has a specific form of asymptote in the limit $r/R \rightarrow 0$ \cite{m91,f95,t09}. The inertial range is assumed to extend to that limit, where the asymptotic PDF is related to the exponent $\tau_m$ via a series of differences between the smoothing dimension $D$ and some fractal dimensions on the $D$-dimensional cut of the space. Since these differences are independent of the value of $D$ \cite{m82}, the exponent $\tau_m$ is independent as well.

For the multifractal PDF of $\ln (\varepsilon_{r_n} / \varepsilon_{r_{n+1}})$, there have been proposed various theories. Some of them lead to $\langle \varepsilon_r^m / \varepsilon_R^m \rangle$ even at $m < 0$ \cite{ms87}. The others do not as in our case of the stable distribution \cite{sl94}.

To any stable distribution that is not Gaussian, the large deviation theory is not applicable. It requires that the tail of the PDF of $\ln (\varepsilon_r / \varepsilon_R) / \ln (r/R)$ decays exponentially as $\ln (r/R) \rightarrow -\infty$, i.e., as $N \rightarrow +\infty$. This is not the case for the PDF of the stable distribution that retains its power-law tail \cite{t09}. Especially for $0 < \alpha < 1$ studied here, $\ln (\varepsilon_r / \varepsilon_R) / \ln (r/R)$ diverges in that limit as $N^{1/\alpha -1}$ (see Sec.~\ref{S3}). The fractal dimensions are hence not available.

The above discussion means that any multifractal PDF of $\ln (\varepsilon_{r_n} / \varepsilon_{r_{n+1}})$ is not stable and is not self-similar. An example is the Poisson distribution \cite{sl94,d94,sw95}. Its parameter is ${\mit\Lambda} > 0$. If $\ln (\varepsilon_{r_1} / \varepsilon_{r_2})$ and $\ln (\varepsilon_{r_2} / \varepsilon_{r_3})$ are related to the parameter values ${\mit\Lambda}_1$ and ${\mit\Lambda}_2$, their sum $\ln (\varepsilon_{r_1} / \varepsilon_{r_3})$ is related to a different parameter value ${\mit\Lambda}_1+{\mit\Lambda}_2$, which results in a different shape of the PDF.

The independence of the scaling exponent $\tau_m$ from the smoothing dimension $D$ is an important characteristic of any multifractal PDF. Among the dimensions $D=1$, $2$, and $3$, the exponent $\tau_m$ is unique. Since it has to satisfy Eq.~(\ref{eq1b}) for $D \ge 1$, it satisfies Eq.~(\ref{eq1c}) at most for $D=1$, although we consider that $\tau_m$ is not unique and satisfies Eq.~(\ref{eq1c}) for each value of $D$ as in our case of the stable distribution.

Multifractality is thus distinct from our log-stability. The underlying assumption for using the large deviation theory is that the intermittency is asymptotically determined in the course of the mean energy transfer to the smaller scales, i.e., the energy cascade \cite{o49}. Nevertheless, the energy is transferred locally and instantaneously to the larger scales as well as to the smaller scales \cite{igk09}. These scales interact with one another and should have settled into a self-similar state that is to be described by a stable distribution of $\ln (\varepsilon_r / \varepsilon_R)$. In addition, it is too idealized to ignore the existence of the dissipation range by taking the limit $r/R \rightarrow 0$ for scales $r$ in the inertial range. If the limit is not taken, the large deviation theory is not of use. The exponent $\tau_m$ remains available from the power-law scaling itself, but it has to be positioned within some alternative to the multifractal framework.

The same discussion applies to other frameworks such as those based on the central limit theorem \cite{y66} and on the {\it generalized} central limit theorem \cite{sl87}, aside from the consistency or inconsistency with Eq.~(\ref{eq1b}) and so on.

With the exponents $\tau_m$ and $\zeta_m$ observed thus far in numerical simulations and in experiments, the multifractal and other theories exhibit a consistency almost comparable to that of our log-stable law \cite{c97,g02,b10}. The theories are not judged definitely at this level of observations. As for the observational judgment between the log-stable law and the others, it is essential to study whether or not the PDF of $\ln (\varepsilon_r / \varepsilon_R)$ is self-similar among scales $r$ in a wide enough inertial range. This is to be expected in the future.

\section{Concluding Remarks} \label{S8}

The small-scale intermittency has been studied about a fully developed state of three-dimensional homogeneous and isotropic turbulence in an incompressible fluid. For the rate $\varepsilon_r$ of energy dissipation smoothed over length scale $r$ in the inertial range up to the largest scale $R$, we have followed Kida \cite{k91} to impose a self-similarity on the PDF of $\ln (\varepsilon_r / \varepsilon_R)$. Such a self-similarity is expected from interactions among the scales through the local and instantaneous transfer of the kinetic energy. The result is an extension of the 1941 theory of Kolmogorov \cite{ko41}, i.e., a one-parameter framework where $\ln (\varepsilon_r / \varepsilon_R)$ obeys some stable distribution. As a statistical framework of the self-similar PDF of $\ln (\varepsilon_r / \varepsilon_R)$, this log-stable law is necessary and sufficient.

The scaling exponents have been obtained for the scaling of the dissipation rate $\langle \varepsilon_r^m \rangle \varpropto r^{\tau_m}$ in Eqs.~(\ref{eq9b}) and (\ref{eq10b}) and for the scaling of the two-point velocity difference $\langle \vert \delta u_r^m \vert \rangle \varpropto r^{\zeta_m}$ in Eq.~(\ref{eq13}). They are consistent with theoretical constraints such as Eqs.~(\ref{eq1b}), (\ref{eq1c}), and (\ref{eq11b}) and with the observed exponents in Fig.~\ref{f2}.

We have attempted to explain the observed value of our parameter $\alpha$. This is also to explain the existence of organized structures in the turbulence because our framework is just a statistical description of them. In fact, the observed value of $\alpha$ is distinct from the value $\alpha \rightarrow 1$ for the classical case of no structures, i.e., the 1941 theory of Kolmogorov \cite{ko41}. By using Eq.~(\ref{eq16}), we have related the observed value of $\alpha$ to the minimum width of the inertial range and thereby to the maximum mean efficiency of the energy dissipation. It follows that the intermittency occurs and the structures are organized so as to maximize the mean efficiency of the dissipation.

We have not considered the fluctuations of $\varepsilon_R$ at the largest scale $R$ of the inertial range. They do not have to be considered if the power-law scaling $\langle \varepsilon_r^m \rangle \varpropto r^{\tau_m}$ is exact. It means that the inertial range is not characterized by any of the scales, including scales of those fluctuations. This is the case of our framework where $\varepsilon_r / \varepsilon_R$ is set to be independent of $\varepsilon_R$, but we would have to study the very case of the actual turbulence. Oboukhov \cite{o62} proposed a PDF of those fluctuations. Since it has turned out to be a good approximation \cite{m09,m13}, it would be useful to that study.

The stable distribution is also expected for a logarithm of any other positive random variable, e.g., enstrophy \cite{igk09,c97}, if its PDF is self-similar in a range of the scales. This is not limited to variables of the turbulence. For example, a power-law scaling is found in a variety of fields \cite{m82} such as the cosmological density field observed through the galaxy number density \cite{tk69}. They have been discussed to be fractal or multifractal, but some of them could be log-stable. It is of interest to search for the stable distribution among logarithms of these and other positive random variables.

\begin{acknowledgments}
This work was supported in part by KAKENHI Grant No. 25340018. The author is grateful to M. Morikawa for interesting discussions.
\end{acknowledgments}


\begin{thebibliography}{999}

\bibitem{ko41} A.\,N. Kolmogorov, Dokl. Akad. Nauk SSSR {\bf 30,} 301 (1941). 

\bibitem{vn49} J. von Neumann, Report to Office of Naval Research, 1949 (unpublished). Reprinted in {\it Collected Works}, edited by A.\,H. Taub (Pergamon, New York, 1963), vol. 6, p. 437.

\bibitem{bt49} G.\,K. Batchelor and A.\,A. Townsend, Proc. R. Soc. London A {\bf 199,} 238 (1949).

\bibitem{gsm62} H.\,L. Grant, R.\,W. Stewart, and A. Moilliet, J. Fluid Mech. {\bf 12, } 241 (1962).

\bibitem{igk09} T. Ishihara, T. Gotoh, and Y. Kaneda, Annu. Rev. Fluid Mech. {\bf 41,} 165 (2009).

\bibitem{n69} E.\,A. Novikov, Soviet Phys. Dokl. {\bf 14,} 104 (1969).

\bibitem{n94} E.\,A. Novikov, Phys. Rev. E {\bf 50,} R3303 (1994).

\bibitem{wf00} T. Watanabe and H. Fujisaka, J. Phys. Soc. Japan {\bf 69,} 1672 (2000).

\bibitem{ko62} A.\,N. Kolmogorov, J. Fluid Mech. {\bf 13,} 82 (1962).

\bibitem{f71} W. Feller, {\it An Introduction to Probability Theory and its Applications}, 2nd ed. (Wiley, New York, 1971), vol. 2.

\bibitem{m82} B.\,B. Mandelbrot, {\it The Fractal Geometry of Nature} (Freeman, San Francisco, 1982).

\bibitem{s13} K.-I. Sato, {\it L\'evy Processes and Infinitely Divisible Distributions}, 2nd ed. (Cambridge University Press, Cambridge, 2013). 

\bibitem{k91} S. Kida, J. Phys. Soc. Japan {\bf 60,} 5 (1991); {\bf 61,} 4671 (1992).  

\bibitem{pnp96} G. Pedrizzetti, E.\,A. Novikov, and A.\,A. Praskovsky, Phys. Rev. E {\bf 53,} 475 (1996).

\bibitem{f91} U. Frisch, Proc. R. Soc. London A {\bf 434,} 89 (1991).

\bibitem{m07} H. Mouri, A. Hori, and Y. Kawashima, Phys. Fluids {\bf 19,} 055101 (2007).

\bibitem{cf94} P. Constantin and C. Fefferman, Nonlinearity {\bf 7,} 41 (1994).

\bibitem{c97} S. Chen, K.\,R. Sreenivasan, and M. Nelkin, Phys. Rev. Lett. {\bf 79,} 1253 (1997).

\bibitem{g02} T. Gotoh, D. Fukayama, and T. Nakano, Phys. Fluids, {\bf 14,} 1065 (2002).

\bibitem{b10} R. Benzi, L. Biferale, R. Fisher, D.\,Q. Lamb, and F. Toschi, J. Fluid Mech. {\bf 653,} 221 (2010).

\bibitem{w96} L.-P. Wang, S. Chen, J.\,G. Brasseur, and J.\,C. Wyngaard, J. Fluid Mech. {\bf 309,} 113 (1996).

\bibitem{ab06} R.\,A. Antonia and P. Burattini, J. Fluid Mech. {\bf 550,} 175 (2006). 

\bibitem{ll59} L.\,D. Landau and E.\,M. Lifshitz, {\it Fluid Mechanics} (Pergamon, London, 1959).

\bibitem{m09} H. Mouri, A. Hori, and M. Takaoka, Phys. Fluids {\bf 21,} 065107 (2009).

\bibitem{ys05} V. Yakhot and K.\,R. Sreenivasan, J. Stat. Phys. {\bf 121,} 823 (2005).

\bibitem{pf85} G. Parisi and U. Frisch, in {\it Turbulence and Predictability in Geophysical Fluid Dynamics}, edited by M. Ghil, R. Benzi, and G. Parisi (North-Holland, Amsterdam, 1985), p. 84.

\bibitem{m91} B.\,B. Mandelbrot, Proc. R. Soc. London A {\bf 434,} 79 (1991).

\bibitem{f95} U. Frisch, {\it Turbulence: The Legacy of A.\,N. Kolmogorov} (Cambridge University Press, Cambridge, 1995).

\bibitem{t09} H. Touchette, Phys. Rep. {\bf 478,} 1 (2009).

\bibitem{ms87} C. Meneveau and K.\,R. Sreenivasan, Phys. Rev. Lett. {\bf 59,} 1424 (1987). 

\bibitem{sl94} Z.-S. She and E. Leveque, Phys. Rev. Lett. {\bf 72,} 336 (1994). 

\bibitem{d94} B. Dubrulle, Phys. Rev. Lett. {\bf 73,} 959 (1994).

\bibitem{sw95} Z.-S. She and E.\,C. Waymire, Phys. Rev. Lett. {\bf 74,} 262 (1995).

\bibitem{o49} L. Onsager, Nuovo Cimento, Suppl. {\bf 6,} 279 (1949).

\bibitem{y66} A.\,M. Yaglom, Soviet Phys. Dokl. {\bf 11,} 26 (1966).

\bibitem{sl87} D. Schertzer and S. Lovejoy, J. Geophys. Res. D {\bf 92,} 9693 (1987). 

\bibitem{o62} A.\,M. Oboukhov, J. Fluid Mech. {\bf 13,} 77 (1962).

\bibitem{m13} H. Mouri, Phys. Rev. E {\bf 88,} 042124 (2013).

\bibitem{tk69} H. Totsuji and T. Kihara, Pub. Astron. Soc. Japan {\bf 21,} 221 (1969).



\end{thebibliography}
\end{document}